\documentclass{aa}
\usepackage{graphicx}
\usepackage{longtable}
\begin{document}

\title{The age of the oldest Open Clusters}

\author{Maurizio Salaris\inst{1,2}, Achim Weiss\inst{2}
\and
Susan M. Percival\inst{1} 
}

\offprints{M. Salaris}

\institute{Astrophysics Research Institute, Liverpool John Moores University,
           Twelve Quays House, Egerton Wharf, Birkenhead, CH41 1LD, UK\\
           \email{ms,smp@astro.livjm.ac.uk}
           \and
Max-Planck-Institut f\"ur Astrophysik, Karl-Schwarzschild-Strasse 1,
85758, Garching, Germany\\
\email{weiss@mpa-garching.mpg.de} 
    }

\date{Received ; accepted}

   \abstract{We determine ages of 71 old Open Clusters by a two-step
   method: we use main-squence fitting to 10 selected clusters, in
   order to obtain their distances, and derive their ages from
   comparison with our own isochrones used before for Globular
   Clusters. We then calibrate the morphological age indicator
   $\delta(V)$, which can be obtained for all remaining clusters, in
   terms of age and metallicity. Particular care is taken to ensure
   consistency in the whole procedure.
   The resulting Open Cluster ages connect well to our previous
   Globular Cluster results. From the Open Cluster sample, as well
   as from the combined sample, questions regarding the formation process
   of Galactic components are addressed. 
   The age of the oldest open clusters (NGC~6791 and Be~17) is of the 
   order of 10~Gyr. We determine a delay by 2.0$\pm$1.5 Gyr between the start 
   of the halo and thin disk formation, whereas 
   thin and thick disk started to form approximately at the 
   same time. We do not find any significant age--metallicity relationship for
   the open cluster sample. The cumulative age distribution of the
   whole open cluster sample shows a moderately significant ($\sim
   2\sigma$ level) departure from
   the predictions for an exponentially declining dissolution
   rate with timescale of 2.5 Gyr. The cumulative age
   distribution does not show any trend with galactocentric distance,
   but the clusters with larger height to the Galactic plane  
   have an excess of objects between 2--4 and 6 Gyr with respect to
   their counterpart closer to the plane of the Galaxy.
   }
\authorrunning{M. Salaris et al.}
\titlerunning{Age of the oldest Open Clusters}
   \maketitle

\section{Introduction}

The theory of the formation of galaxies is without any doubt one of
the outstanding problems of astrophysics. Although in the past decades
considerable progress has been made, we do not have yet a complete
and definitive picture of how galaxies form. 
As discussed by, e.g., Freeman \& Bland-Hawthorn~(2002), 
a detailed study of the formation of the Galaxy lies at the core of
understanding the complex processes leading to the formation of
external galaxy systems.
A way to shed some light on this problem is to study the timescale for
the formation of the different Galactic populations, e.g., halo, thick
disk, thin disk
and bulge, by means of stellar age dating.
The most reliable stellar ages are obtained for the star clusters belonging
to the various populations, i.e., the globular clusters (GCs) in the
halo, thick disk and bulge, and the open clusters (OCs) in the thin
disk. The advantage of dating star clusters over individual stars --
whose age determination relies entirely on the knowledge of
individual metallicities, effective temperatures and gravities (or absolute
magnitudes), which have to be fitted by the appropriate theoretical
model --  stems from the fact that star clusters are made of coeval objects, largely with 
the same initial chemical composition and located at the same
distance, so that it is possible to use morphological parameters deduced
from theoretical isochrones in order to derive their age. In this way 
one can bypass the thorny problem of determining a reliable empirical and
theoretical temperature scale, and of acquiring high resolution spectroscopy for large
samples of stars.   

In a series of papers published in the last 6 years (Salaris et
al.~1997; Salaris \& Weiss~1997, 1998; Salaris \&
Weiss~2002, hereinafter SW02), 
we have addressed the problem of the timescale for the formation of the
halo and thick disk by homogeneously determining the age of a 
large sample of Galactic GCs.
The latest SW02 study (including 55 GCs) concluded that metal poor
clusters (up to [Fe/H] between $-$1.6 and $-$1.2,
depending on the adopted [Fe/H] scale) are coeval within $\sim$1~Gyr,
with an age of the order of 12--13~Gyr, whereas the more metal rich ones
show an age spread, are on average younger and display a weak age-metallicity
relationship (age decreasing with increasing [Fe/H]). This result is
in agreement with other independent analyses, such as that by Rosenberg et al.~(1999).
When searching for relationships between age and position within the halo,
it was found that the age spread starts from galactocentric distances
($R_\mathrm{gc}$) between 8 and 13
kpc outwards, the precise value depending again on the
adopted [Fe/H] scale.

It is now important to address the question of when the thin disk
started to build up, relative to the thick disk and halo.
This can be accomplished by studying the age distribution of the
oldest OCs. In general, OCs are expected to be disrupted easily by
encounters with massive clouds in the disk (Spitzer~1958); 
however, the most massive OCs or those with orbits
that keep them far away from the Galactic plane for most of their lifetimes
are expected to survive for longer periods of time.
These old objects are therefore test particles -- in analogy to the GCs --
probing the earliest stages of the formation of the disk.
It is essential to determine their ages 
using stellar models and methods which place them on the same
scale as GC ages. An analysis of this kind, based on homogeneous age
dating of all the known old OCs and a large sample of GCs,
employing the latest generation of stellar
models is still lacking (see, e.g., Liu \&
Chaboyer~2000 for a study of this kind, but considering only a
very small number of OCs and GCs), and this paper is intended 
to fill this gap.

Here we will reanalyze the old OC sample reviewed by
Friel~(1995, hereafter F95),
and based on the seminal papers by Phelps, Janes \& Montgomery~(1994) and Janes \&
Phelps~(1994, hereafter JP94), to which we have added two additional clusters
(ESO~093-SC08 and vdBH~176) recently studied 
by Phelps \& Schick~(2003). This should contain approximately 
all presently known old OCs.
Our aim is to determine their age on a
scale consistent with the GC ages determined by SW02,
to study the existence of possible relationships between age, position within the disk and
[Fe/H], and to compare their ages with the GC population.
In Sect.~2 we describe the cluster sample and the techniques used to
determine their age. The resulting age distribution is analyzed in
Sect.~3, while Sect.~4 deals with the comparison with the GC ages
by SW02. A summary and conclusions follow in Sect.~5.

\section{Cluster sample and age determination method}

We consider a total of 71 clusters -- 69  
from Friel~(1995) and 2 from Phelps \& Schick~(2003) -- 
whose morphological age parameter $\delta(V)$ is equal to or larger than the
value for Praesepe, i.e. $\delta(V)$=0.3 (see below for the definition of
$\delta(V)$). The definition of old OCs by 
JP94 is slightly different: they considered
as `old' all OCs where $\delta(V)$ is larger than zero. We have not 
included objects with $\delta(V)<$0.3, because we did not have clusters
with $\delta(V)$ between 0 and 0.3 that could be used to calibrate adequately 
a relationship between this parameter and cluster age (see below for
details about the calibration). 

Friel's~(1995) sample is mainly the same as the one studied
by JP94, with only a few additions. 
JP94 have discussed in detail the completeness of their sample of old 
OCs, and concluded
that most probably the number of undetected old clusters is small and
should not have a major effect on the overall age distribution, 
even though the properties of the age distribution perpendicular to the
Galactic plane may be affected by still undetected old OCs, which
should be preferentially located very close to the plane of the Galaxy.

The [Fe/H] values for our sample are taken whenever possible (38
clusters) from Gratton~(2000, hereafter G00), who transformed various metallicity
scales based on low resolution spectroscopy onto an homogeneous scale
tied to high resolution [Fe/H] determinations. One exception is the
cluster Praesepe, for which we have employed the Hyades
metallicity (G00 reports a lower value), based on the discussion 
and references in Percival et al.~(2002, hereafter PSK02).
In case of clusters not listed by G00
we have either used the value provided by the WEBDA OC database
(http://obswww.unige.ch/webda/, see Mermilliod~1992) 
when available, to which we attached an error of
0.15 dex (11 clusters), or we assumed [Fe/H]=0.0 with an error of 0.20
dex (22 clusters).
The cluster galactocentric distances and
heights to the Galactic plane are taken from Friel~(1995) and Phelps \&
Schick~(2003). For many of the clusters in our analysis the existing
photometry and/or uncertainties in the cluster parameters 
do not allow to perform a more accurate and homogeneous distance 
determination. Therefore we used the results presented in the
mentioned papers, where more details about this issue can be found.

  \begin{figure}
     \resizebox{\hsize}{!}{\includegraphics{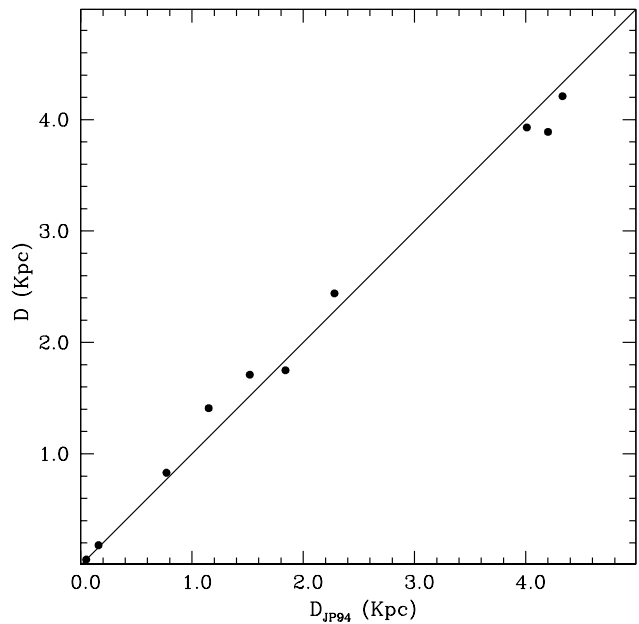}}
      \caption{Comparison between the distances given by JP94 and the
     MS fitting distances we obtain for a subsample of 10 clusters
     (see text for details). The solid line represent the 1:1
     relationship between the two sets of distances. 
              }
         \label{distcompOC}
   \end{figure}

As a test we compared in Fig.~\ref{distcompOC} the distances
we obtained for a subsample of 10 clusters (see next subsection for details) from the
Main Sequence (hereinafter MS) fitting technique, with the JP94 results. We did not
find any statistically significant trend of the difference between the
two sets of distances with respect to our MS fitting
determinations. The mean value of the difference is equal to only 9 pc, with a
dispersion of 160 pc around the mean.

The data for the complete cluster sample are summarized in Table~1; the flag 
attached to each cluster provides the source for the metallicity; a
value equal to 0 or 1 means [Fe/H] from G00, whereas a value of 2
means that the source is the WEBDA database or that there is no
available [Fe/H] determination. A flag equal to 0 denotes the
subsample of clusters that are used for our age calibration, as
explained in the next subsections.

We have used as age indicator the morphological parameter $\delta(V)$
defined by JP94, which is similar to the $\Delta(V)$ parameter used in
GC dating (e.g. SW02), calibrated in terms of
absolute age and
[Fe/H] following the same kind of approach as in JP94, and in Carraro \&
Chiosi~(1994a) for their analysis of a sample of 36 old OCs.

\subsection{The morphological age index $\delta(V)$}

The use of morphological indices that quantify differences in the
Colour-Magnitude Diagram (CMD) of clusters in terms of age differences
is a well established technique (see, e.g., Anthony-Twarog \&
Twarog~1985 and JP94 for OCs; Rosenberg et al.~1999 for GCs); it
allows to establish a relative age ranking among a given cluster
sample, bypassing the well known difficulties with isochrone fitting methods
(see, e.g., VandenBerg et al.~1990;
Sarajedini \& Demarque~1990; Salaris \& Weiss~1997, SW02).

Phelps et al.~(1994) and JP94 have defined two morphological 
parameters, called $\tilde{\delta}(V)$\footnote{In the quoted paper 
this quantity is denoted as $\delta(V)$} and
$\delta1$ which they applied to their sample of old OCs. 
$\tilde{\delta}(V)$ is defined as the magnitude difference between
the cluster turn-off region and the He-burning clump stars. More
precisely, the reference point in the turn-off region is taken as the
inflection point between the turn-off and the base of the giant
branch. This point is well defined and unaffected by the presence of
a binary sequence and/or field stars, according to Phelps et al.~(1994).
$\delta1$ is the difference in colour index between the bluest point on
the MS at the luminosity of the turn-off and the colour of the giant
branch one magnitude brighter than the turn off luminosity. 
A simple linear relationship between $\tilde{\delta}(V)$ and $\delta1$
was found by JP94 when analysing the clusters where {\em both} indices
could be measured, and this was applied by the same authors to obtain an estimate for
$\tilde{\delta}(V)$ for 
those clusters where clump stars were not identified. This estimate we
will call $\delta(V)$ in the following. 
JP94 choose $\delta(V)$ as the primary age indicator, so that for
clusters without visible clump the estimated value described above was
used; for the other clusters the final $\delta(V)$ given by JP94 is the average
between the observed $\tilde{\delta}(V)$ and the one computed from the 
$\delta1-\delta(V)$ conversion described above.

The $\delta(V)$ data  
for our cluster sample are reported in Table~1;
the associated errors are derived from the quality grade assigned
by JP94. Following JP94 we considered errors by, respectively, 0.05 mag for clusters
graded ``a'', 0.15 mag for clusters graded ``b'' and 0.25 mag for
clusters graded ``c''.
In case of the two clusters from Phelps \& Schick~(2003) we have 
assumed an error by 0.25 mag.

\subsection{Calibration of $\delta(V)$}

The cluster $\delta(V)$ values given in Table~1 can be translated into absolute ages by
determining a relationship $\delta(V)$--$t$--[Fe/H], based on a subsample
of clusters with high quality CMDs, spanning the entire [Fe/H] and
$\delta(V)$ range of the full cluster sample, and for which the age can be 
determined with confidence. JP94 determined a relationship between
$\delta(V)$ and age -- neglecting the effect of metallicity -- based
on a sample of OCs and GCs with age determinations obtained by various
authors and with a variety of methods and stellar models. In case of
multiple age determinations for the same objects JP94 averaged the
results from the various authors. As clearly stated by JP94, due to
the heterogeneity of the calibration material, their calibration
was mainly aimed at producing the ranking of the clusters in terms of
relative ages.

Here we wish to obtain a new highly homogeneous and reliable
calibration in terms of absolute ages, based on a
consistent set of updated stellar models. 
We have considered a subsample of 10 OCs (clusters with the flag value
equal to 0 in Table~1)  plus 1 GC (47~Tuc), whose ages have been
determined by fitting the CMD turn-off luminosity with
theoretical isochrones, after determining their distance from 
an empirical MS fitting technique which employs large samples of field
MS stars with accurate $Hipparcos$ parallaxes.
In this way the ages we obtain are firmly
tied to the $Hipparcos$ distance scale.

The stellar models used to determine the OC ages have been
computed with exactly the same updated physics employed for 
calculating the GC isochrones
by Salaris \& Weiss~(1998); these GC isochrones have been used to determine the
ages of the large GC sample analyzed by SW02, and  
we refer the reader to Salaris \& Weiss~(1998) 
for details about the model input physics.
The turn-off stars in the younger clusters in our sample do have
convective cores, and therefore 
we have included in our models overshooting beyond the 
formal boundary of the convective core
(i.e., instantaneous mixing and radiative temperature gradient 
in the overshooting region beyond the boundary of the
convective core fixed by the Schwarzschild criterion),
with an extension 
of 0.2 pressure scale heights 
for masses above 
1.4$M_{\odot}$, and linearly decreasing to zero from 
M=1.4$M_{\odot}$ to M=1.0$M_{\odot}$.
This prescription is in broad agreement with the results obtained by 
Ribas et al.~(2000) from the comparison of stellar models with
eclipsing binary systems,
at least in the mass interval spanned by the stars evolving in the
turn-off region of our OC sample.
We have computed stellar models and isochrones 
for the appropriate metallicity (scaled solar Grevesse \& Noels~1993 heavy
element distribution) of each
of our calibrating clusters, using an He-mass fraction $Y$ that follows
the relationship  $Y=0.248+1.44\,Z$. The primordial He is derived from
the recent results of analyses of the CMB power spectrum 
(see, e.g., the discussion in Cassisi et al.~2003), 
whereas the slope $\Delta(Y)/\Delta(Z)=1.44$
arises from the constraint imposed by the initial He-abundance of 
the standard solar model. In case of 47~Tuc and the other GC ages
discussed in Sect.~4 we have computed selected $\alpha$-enhanced
isochrones (the same metal distribution as in Salaris \& Weiss~1998) at
various metallicities and with the same $Y$--$Z$ relationship as for the
OCs, in order to determine the age of 47~Tuc and revise the ages of
the GCs in SW02 which were computed using $Y=0.230+3.0 \, Z$.

The MS fitting distances to the calibrating OCs are 
from PSK02 and Percival \& Salaris~(2003, hereafter PS03), 
with the exception of NGC~6791 (see below), whilst the distance to 47~Tuc is from
Percival et al.~(2001, hereafter P01); they are all based on two large samples of 
unevolved field MS stars with accurate $Hipparcos$
parallaxes and individual metallicity determinations. The more metal
rich sample of 54 dwarfs with
[Fe/H] between $\sim -0.4$ and $\sim +0.3$ ({\em field dwarfs}) has been used to derive
the distances to the calibrating OCs, while the distance to 47~Tuc
has been obtained using a sample of 43 more metal-poor dwarfs with 
[Fe/H] between $\sim -1.0$ and $\sim -0.3$ ({\em subdwarfs}).
Table~2 contains the results from the MS fitting 
distance determinations; reddenings are from Twarog et al.~(1997) with
an associated error of $\pm$0.02 mag as adopted by Sarajedini~(1999).
It is important to notice that, as discussed in PSK02, 
our MS fitting distances to the
Hyades and Praesepe agree well with the $Hipparcos$
parallax measurements.

A detailed description of the distance determination method 
is given in the three papers mentioned above, together with the 
sources for the adopted MS CMDs. We just recall here that the method is
based on constructing an empirical template MS 
from the field stars, by applying colour shifts to the individual objects
to account for the differences in metallicity between the field stars
and the cluster. The template is then shifted in magnitude to match
the dereddened and extinction-corrected cluster MS, the
extent of the shift being equal to the distance modulus $(m-M)_{0}$.
The colour shifts for the field star sample have been derived
empirically as discussed in PSK02 and,
strictly speaking, they are applicable only within the metallicity 
range spanned by the field dwarfs themselves, i.e., up to [Fe/H]$\sim$0.3.
In case of the subdwarf sample used for 47~Tuc, the shifts have been
obtained from the differential use of the isochrones by Salaris \&
Weiss~(1998) as discussed in P01.
Whenever possible we have derived MS fitting distances using both the
$(B-V)$ and $(V-I)$ colours, always finding agreement (within the
associated error bars, which, for given reddening and [Fe/H] values,
are typically of a few hundredths of magnitude, and usually the same for both colours) 
between the two values. In this case the 
final distance is the unweighted mean
between the results in $(B-V)$ and in $(V-I)$. The final error budget 
takes into account the errors in the cluster reddening and [Fe/H]
quoted in Tables~1 and 2.

P01 and PSK02 discussed at length the consistency between the 
metallicity scale of the field stars and clusters 
for each of the two separate samples, a necessary prerequisite for the
reliability of the MS fitting distances.
What matters most in our case is the consistency between 
the distances obtained separately for 47~Tuc with the sub\-dwarf sample, 
and for the other
calibrating OCs with the field star sample. In fact, they have been 
determined from two different
samples of field MS stars, with metallicity scales determined independently,
and using different methods to determine the colour shifts.
Consistency between the two sets of distances means
that one should be able to use
the subdwarf sample to derive the MS fitting distance modulus 
to a cluster like, e.g., NGC~2420, which is at the lower metallicity end of the
calibrating OCs,
and recover the value obtained employing the field dwarf sample 
(11.94$\pm$0.07 mag).
Since we get consistent distances in both $(B-V)$ and $(V-I)$
for the distance to 47~Tuc and to other calibrating OCs,
we performed this test with the $(V-I)$ colour only. If consistency
is achieved for this colour, it is automatically ensured for $(B-V)$, too.

  \begin{figure}
     \resizebox{\hsize}{!}{\includegraphics{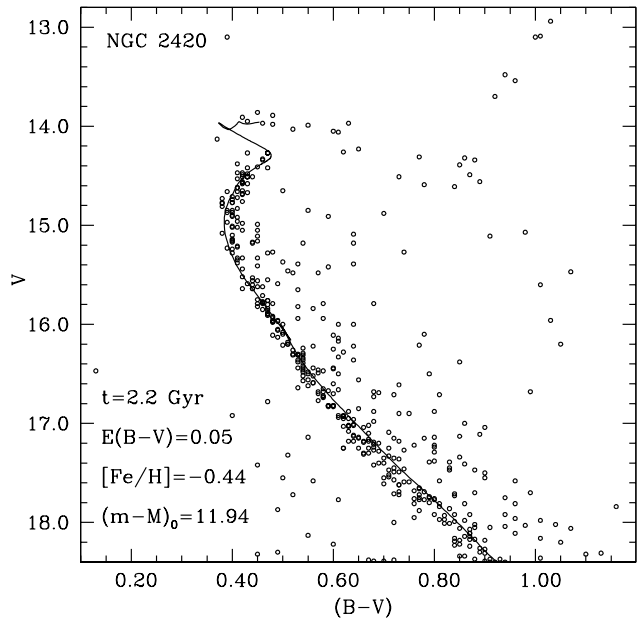}}
      \caption{Best-fitting isochrone for the MS and turn-off region
     of the cluster NGC~2420. Reddening, distance modulus, 
     age and metallicity employed in the fit are given.
      }
         \label{fitage}
   \end{figure}

We performed this test using the entire subdwarf sample and
determining the appropriate colour shifts
from the theoretical isochrones discussed before. We finally obtained a
distance modulus that is within 0.02 mag of the value obtained from the field
dwarf sample, confirming the consistency between the distances
obtained from the two separate samples of subdwarfs and field dwarfs.

We also included NGC~6791 in the calibrating sample, in
order to extend our calibration to very high metallicities, and have
another extremely 'old' (e.g. JP94) calibrating cluster in addition to 47~Tuc.
We used $E(B-V)=0.15\pm0.02$ (Twarog et al.~1997),
[Fe/H]=0.40$\pm$0.06 (G00), and employed the same field dwarf sample
and method as for the other OCs; the cluster MS CMD is 
from the new
photometry by Stetson et al.~(2003). 
We determined the
distance modulus using both the $(B-V)$ and
$(V-I)$ colours (they provide the same distance modulus within 0.01 mag), 
obtaining $(m-M)_0$=12.96$\pm$0.10, as 
reported in Table~2. 

As a note of caution we notice that the cluster NGC~6791 metallicity
is slightly above the upper limit of the [Fe/H] range where our MS 
fitting method is applicable, so that we had to slightly extrapolate
the empirical colour shifts applied to the field dwarfs. However, we
have tested that the exclusion of NGC~6791 from the calibrating sample
does not alter substantially the calibration of the sought
$\delta(V)$-t-[Fe/H] relationship and therefore we retained NGC~6791
in our calibrating sample.

After the MS fitting distances have been determined,
cluster ages for the calibrating clusters have been obtained from
isochrone fitting to the CMD turn off region, 
and are given in Table~1. Figure~\ref{fitage}
shows an example (the cluster NGC~2420) of our age determination.
The error bar for the age includes in quadrature the contributions due
to the uncertainty in the cluster distance modulus and metallicity.

  \begin{figure}
     \resizebox{\hsize}{!}{\includegraphics{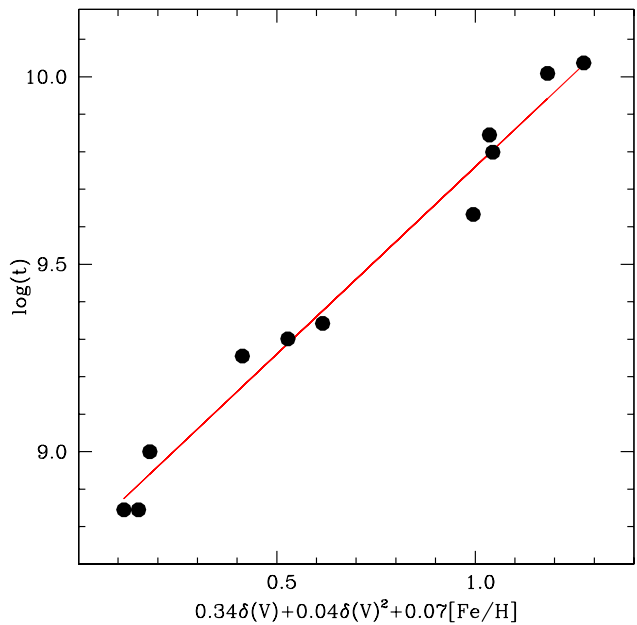}}
      \caption{Fit to the ages of the calibrating clusters (filled
     circles) given in Table~1.
     }
         \label{calibage}
   \end{figure}

We notice that our ages are very similar to the values obtained by
PS03 using the same distance moduli but the
isochrones by Girardi et al.~(2000). We also, as a test, 
determined the ages of our calibrating clusters by means of the 
Lejeune \& Schaerer~(2001) isochrones, obtaining  ages 
within less than 10\% of the values given in Table~1.

With ages, [Fe/H] (all on the G00 scale) and $\delta(V)$ 
values of the calibrating clusters
we determined the sought calibration of age as a function 
of $\delta(V)$ and metallicity.
We found a simple relationship (shown in Fig.~\ref{calibage}) 
between the logarithm of the
cluster age and both [Fe/H] and $\delta(V)$, given by 

\begin{equation} 
{\log(t)}= 0.04 \ \delta(V)^2 + 0.34 \ \delta(V) + 0.07 \ {\rm [Fe/H]} + 8.76 
\label{cal1}
\end{equation} 
with a 1$\sigma$ dispersion equal to 0.062 dex.

The dependence of the logarithm of the age on  $\delta(V)$ 
is not very different from the calibration by JP94, who found ${\log(t)}$ 
to be proportional to $0.256 \, \delta(V) + 0.0662 \,
\delta(V)^2$, without including a metallicity term.
Interestingly, Carraro \& Chiosi~(1994a) calibrated ${\log(t)}$ in terms of 
a morphological parameter similar to $\delta(V)$, and determined a
dependence on metallicity equal to $0.08 \ {\rm [Fe/H]}$, almost
identical to our result for the JP94 $\delta(V)$. 

It is interesting to notice that the [Fe/H] dependence of Eq.~(1)
is qualitatively the same as in theoretical models. In fact, by
using, e.g., the Girardi et al.~(2000) isochrones, we have 
computed the magnitude difference between He-burning clump and turn
off for the age and [Fe/H] range of the studied clusters. 
We found that, for a given value of this magnitude difference, 
metal poorer isochrones provide lower
ages, as predicted by Eq.~(1). The reason is that in
this [Fe/H] range, once the age is fixed, a decrease of the metallicity
increases the clump luminosity more than the turn off one. 
Owing to the fact that the clump level is weakly dependent on age for
our relevant age range, it is therefore clear that 
isochrones with lower metallicity have to be younger in order 
to show the same clump-turn off magnitude 
difference as more metal rich ones. The opposite happens in the regime
of metal poor globular clusters,
where the dependence of the horizontal branch magnitude on
metallicity is weaker than the turn off one (at fixed age).


\section{The age of the old OCs}

In order to determine the ages of the remaining 61 OCs in our sample 
we have applied Eq.~(1) to their $\delta(V)$ and [Fe/H] values
displayed in Table~1.

   \begin{figure}
     \resizebox{\hsize}{!}{\includegraphics{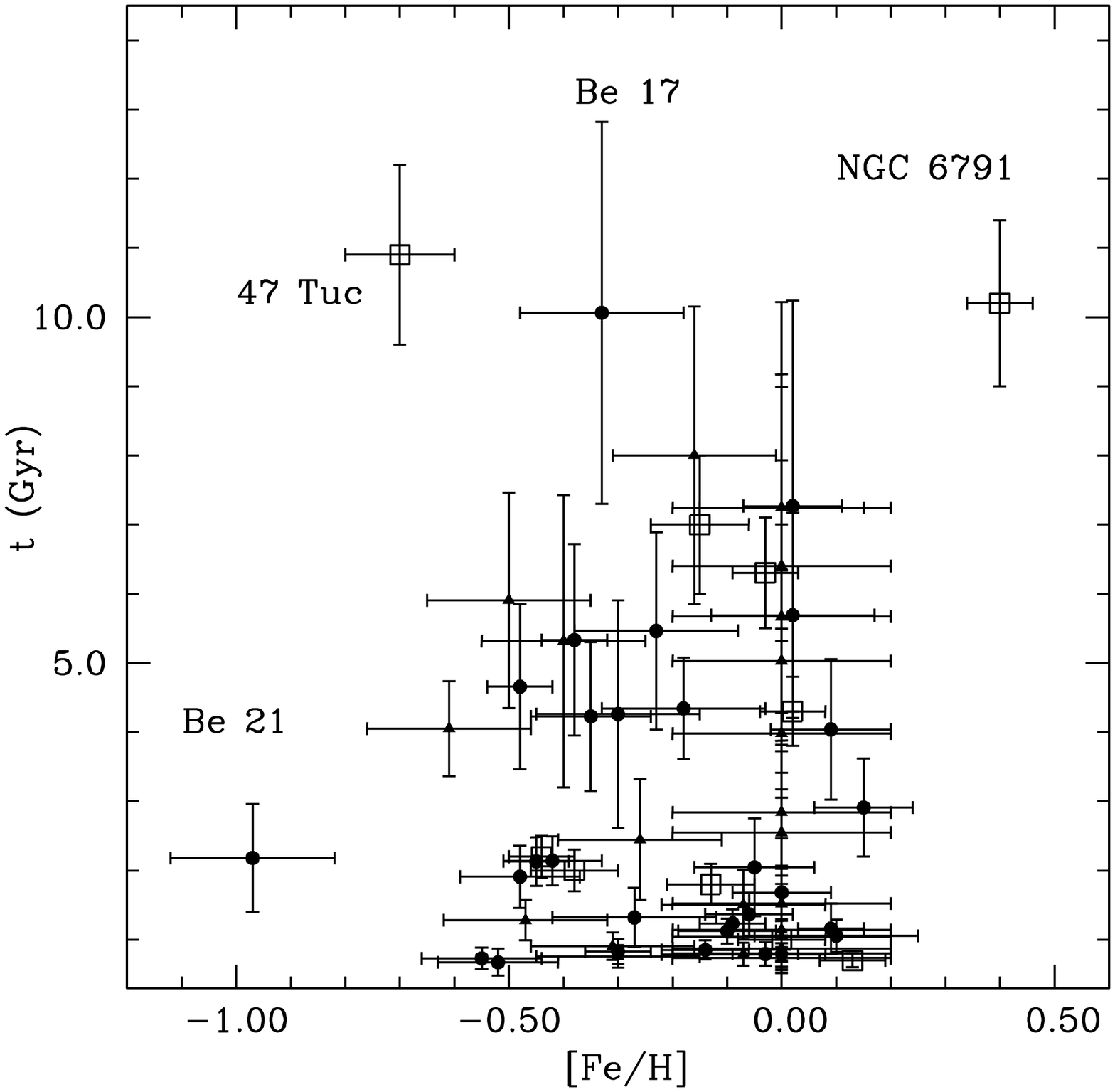}}
      \caption{Ages for the 71 old OCs plus 47~Tuc,
       as a function of the cluster [Fe/H]. Open squares denote the 11
       calibrating clusters. Individual clusters discussed in the text
       are labelled.
       }
         \label{ageOC}
   \end{figure}


\begin{table*} 
\renewcommand{\arraystretch}{0.80} 
\caption[]{Cluster data. The columns display, respectively, 
cluster name, value of the $\delta$(V) morphological
parameter and its associated error, [Fe/H] and associated error, age in Gyr and
associated error, galactocentric distance in kpc, height to the
Galactic plane in pc, source of [Fe/H] value, previous age estimate on
the JP94 scale (see text for details). The last 11
clusters are the calibrating clusters for our t--[Fe/H]--$\delta$(V) relationship.}
\begin{minipage}{\textwidth} 
\begin{tabular}{lccrcrrrrcc} \hline 
            Cluster
            & $\delta$(V) 
            & $\sigma(\delta$(V))
            & [Fe/H]
            & $\sigma$([Fe/H])
            & t (Gyr)
            & $\sigma$(t)
            & $R_\mathrm{gc}$ (kpc)
            & $z$ (pc)
            & flag
            & $\rm t_{JP94}$ (Gyr)\\
\hline

King~2      &  2.2  &  0.15 &   0.00 &   0.20 &   5.03   &   1.31  &    12.98  &   $-$510 &  2 &  5.6\\
IC~166      &  1    &  0.25 & $-$0.27 &   0.15 &  1.32  &    0.43  &    10.74  &   $-$10  &  1 &  1.5\\
NGC~752     &  0.9  &  0.05 & $-$0.09 &   0.06 &  1.24  &    0.20  &     8.75  &   $-$145 &  1 &  1.4\\
Be~66       &  2.0  &  0.25 &    0.00 &   0.20 &  3.98  &     1.52 &    12.59  &      20 &  2 &  4.4\\
NGC~1193    &  2.1  &  0.15 & $-$0.35 &   0.11 &  4.23  &     1.08 &    12.00  &   $-$845 &  1 &  4.9\\
King~5      &  0.4  &  0.15 & $-$0.30 &   0.15 &  0.76  &    0.16  &   10.34   &   $-$163  & 2  & 0.9\\
NGC~1245    &  0.7  &  0.15 &   0.10 &   0.15 &  1.06   &   0.23   &   11.09   &   $-$465  & 1  & 1.0\\
NGC~1798    &  1.0  &  0.15 & $-$0.47 &   0.15 &  1.28  &    0.29  &    11.79  &     290 &  2 &  1.5\\
NGC~1817    &  0.8  &  0.05 & $-$0.10 &   0.09 &  1.12  &     0.18 &    10.26  &   $-$410 &  1 &  1.3\\
Be~17       &  2.8  &  0.15 & $-$0.33 &   0.15 &  10.06  &    2.77 &    10.89  &   $-$155 &  1 & 12.6\\
Be~18       &  2.3  &  0.15 &   0.02 &   0.15 &  5.69   &    1.49  &    12.09  &     325 &  1 &  5.6\\
Be~20       &  2.1  &  0.05 & $-$0.61 &   0.15 &  4.05  &    0.69  &    16.12  &   $-$2420&  2 &  4.9\\
Be~21       &  1.6  &  0.25 & $-$0.97 &   0.15 &  2.18  &    0.78  &    14.27  &   $-$255 &  1 &  2.8\\
Be~22       &  2.1  &  0.25 & $-$0.30 &   0.15 &  4.26  &     1.65 &    11.92  &   $-$530 &  1 &  3.5\\
NGC~2141    &  1.6  &  0.25 & $-$0.26 &   0.15 &  2.45  &     0.88 &    12.60  &   $-$430 &  2 &  2.8\\
NGC~2158    &  1.4  &  0.15 & $-$0.48 &   0.11 &  1.91  &    0.45  &    12.36  &     120 &  1 &  2.2\\
NGC~2194    &  0.5  &  0.15 &   0.00 &   0.20 &  0.87   &   0.19   &    11.06  &   $-$110 &  2 &  1.0\\
NGC~2192    &  0.6  &  0.15 & $-$0.31 &   0.15 &  0.91  &     0.20 &    11.88  &     635 &  2 &  1.1\\
NGC~2236    &  0.4  &  0.25 &   0.00 &   0.20 &  0.80   &     0.24 &    11.61  &   $-$100 &  2 &  0.9\\
NGC~2243    &  2.2  &  0.15 & $-$0.48 &   0.06 &  4.66  &     1.20 &    10.76  &   $-$1130&  1 &  5.6\\
Tr~5        &  2.3  &  0.25 &   0.00 &   0.20 &  5.67   &    2.26  &    11.13  &     50  &  2 &  4.9\\
NGC~2266    &  0.5  &  0.25 &   0.00 &   0.20 &  0.87   &   0.26   &    11.80  &     600 &  2 &  1.0\\
Be~29       &  2.1  &  0.05 & $-$0.18 &   0.15 &  4.34  &    0.74  &    18.72  &     1465&  1 &  5.6\\
Be~31       &  2.3  &  0.25 & $-$0.40 &   0.15 &  5.32  &     2.11 &    12.02  &     340 &  2 &  3.5\\
Be~30       &  0.3  &  0.15 &   0.00 &   0.20 &  0.74  &   0.16   &    10.58  &     120 &  2 &  0.9\\
Be~32       &  2.4  &  0.15 & $-$0.50 &   0.15 &  5.91  &     1.56 &    11.30  &     235 &  2 &  7.2\\
To~2        &  1.5  &  0.05 & $-$0.45 &   0.06 &  2.13  &     0.35 &    13.08  &   $-$725 &  1 &  2.5\\
NGC~2324    &  0.3  &  0.25 & $-$0.52 &   0.11 &  0.67  &    0.20  &    11.29  &     185 &  1 &  0.9\\
NGC~2354    &  0.8  &  0.25 &   0.00 &   0.20 &  1.14   &   0.36   &     9.56  &   $-$215 &  2 &  1.3\\
NGC~2355    &  0.4  &  0.15 & $-$0.07 &   0.15 &  0.79  &    0.17  &    10.52  &     450 &  2 &  0.9\\
NGC~2360    &  0.5  &  0.05 & $-$0.14 &   0.08 &  0.85  &    0.14  &     9.28  &   $-$30  &  1 &  1.0\\
Haf~6       &  0.3  &  0.25 &   0.00 &   0.20 &  0.73   &   0.21   &    10.91  &     15  &  2 &  0.9\\
Me~66       &  2.3  &  0.15 & $-$0.38 &   0.06 &  5.33  &     1.38 &     9.44  &   $-$710 &  1 &  6.3\\
Me~71       &  0.5  &  0.15 & $-$0.30 &   0.06 &  0.83  &    0.18  &    10.46  &     210 &  1 &  1.0\\
AM~2        &  2.5  &  0.15 &   0.00 &   0.15 &  7.24   &   1.93   &    14.06  &   $-$740 &  2 &  8.3\\
NGC~2506    &  1.5  &  0.05 & $-$0.42 &   0.09 &  2.14  &    0.35  &    10.81  &     555 &  1 &  2.5\\
Pismis~2    &  1.1  &  0.25 & $-$0.07 &   0.15 &  1.51  &    0.50  &     9.47  &   $-$165 &  2 &  1.7\\
Pismis~3    &  1.7  &  0.25 &   0.00 &   0.20 &  2.84   &   1.04   &     8.83  &     10  &  2 &  3.1\\
NGC~2627    &  1.6  &  0.15 &   0.00 &   0.20 &  2.55   &   0.62   &     9.28  &     220 &  2 &  2.8\\
NGC~2660    &  0.4  &  0.15 & $-$0.55 &   0.11 &  0.73  &    0.16  &     9.18  &   $-$155 &  1 &  0.9\\
NGC~2849    &  0.5  &  0.25 &   0.00 &   0.20 &  0.87   &   0.26   &    10.64  &     630 &  2 &  1.0\\
092-SC18    &  2.2  &  0.25 &   0.00 &   0.20 &  5.03   &    1.98  &     9.00  &   $-$740 &  2 &  5.6\\
NGC~3680    &  1    &  0.15 &   0.06 &   0.08 &  1.37   &   0.31   &     8.27  &     310 &  1 &  1.5\\
Cr~261      &  2.6  &  0.15 & $-$0.16 &   0.15 &  8.00   &    2.16  &     7.49  &   $-$250 &  2 &  9.5\\
NGC~4815    &  1.1  &  0.25 &   0.00 &   0.20 &  1.52   &    0.51  &     7.90  &    $-$80 &  2 &  1.7\\
NGC~5822    &  0.8  &  0.25 &   0.09 &   0.06 &  1.16   &   0.36   &     7.94  &      45 &  1 &  1.3\\
\hline 				
\label{OCdata} 
\end{tabular} 
\end{minipage} 
\end{table*} 

\setcounter{table}{0}
\begin{table*} 
\renewcommand{\arraystretch}{0.80} 
\caption[]{contd.}
\begin{minipage}{\textwidth} 
\begin{tabular}{lccrcrrrrcc} \hline 
            Cluster
            & $\delta$(V) 
            & $\sigma(\delta$(V))
            & [Fe/H]
            & $\sigma$([Fe/H])
            & t (Gyr)
            & $\sigma$(t)
            & $R_\mathrm{gc}$ (kpc)
            & $z$ (pc)
            & flag
            & $\rm t_{JP94}$ (Gyr)\\
\hline

IC~4651     &  1.2  &  0.15 &   0.00 &   0.09 &  1.68   &    0.39  &     7.65  &    $-$125 &  1 &  1.8\\
IC~4756     &  0.4  &  0.15 & $-$0.03 &   0.06 &  0.79  &    0.17  &     8.19  &      35 &  1 &  0.9\\
Be~42       &  0.4  &  0.25 &   0.00 &   0.20 &  0.80   &     0.24 &     7.60  &    $-$45 &  2 &  0.9\\
NGC~6802    &  0.4  &  0.25 &   0.00 &   0.20 &  0.80   &     0.24 &     7.96  &      15 &  2 &  0.9\\
NGC~6819    &  1.7  &  0.15 &   0.15 &   0.09 &  2.91   &   0.71   &     8.18  &     300 &  1 &  3.1\\
NGC~6827    &  0.5  &  0.25 &   0.00 &   0.20 &  0.87   &   0.26   &     8.32  &   $-$355 &  2 &  1.0\\
NGC~6939    &  1.4  &  0.25 & $-$0.05 &   0.11 &  2.05  &    0.71  &     8.70  &     255 &  1 &  2.2\\
Be~54       &  2.5  &  0.25 &   0.02 &   0.09 &  7.27   &    2.97  &     8.54  &   $-$165 &  1 &  7.2\\
NGC~7044    &  0.7  &  0.15 &   0.00 &   0.20 &  1.04   &   0.23   &     9.08  &   $-$280 &  2 &  1.2\\
Be~56       &  2.3  &  0.25 &   0.00 &   0.20 &  5.67   &    2.26  &     9.92  &   $-$515 &  2 &  6.3\\
NGC~7142    &  2    &  0.15 &   0.09 &   0.11 &  4.04   &   1.02   &     9.70  &     485 &  1 &  4.4\\
King~9      &  2    &  0.25 &   0.00 &   0.20 &  3.98   &    1.52  &    10.41  &   $-$145 &  2 &  4.4\\
King~11     &  2.3  &  0.15 & $-$0.23 &   0.15 &  5.46  &    1.43  &     9.69  &     245 &  1 &  6.3\\
093-SC08    &  2.4  &  0.25 &   0.00 &   0.20 &  6.40   &    2.59  &    13.00  &   $-$1000 &  2 &  7.3\\
vdBH~176    &  2.5  &  0.25 &   0.00 &   0.20 &  7.24   &    2.98  &    12.00  &    1350 &  2 &  8.6\\ 
\hline 		
Calibrating clusters & & & & & & & & & \\
\hline
M~67        &  2.3  &  0.05 &   0.02 &   0.06 &  4.30   &   0.50   &     9.05  &     405 &  0 &  6.3\\
NGC~2477    &  0.5  &  0.15 &   0.00 &   0.08 &  1.00   &   0.30   &     8.89  &   $-$115 &  0 &  1.0\\
NGC~188     &  2.4  &  0.15 & $-$0.03 &   0.06 &  6.30  &     0.80 &     9.35  &     580 &  0 &  7.2\\
NGC~7789    &  1.1  &  0.05 & $-$0.13 &   0.08 &  1.80  &    0.30  &     9.44  &   $-$170 &  0 &  1.7\\
Be~39       &  2.4  &  0.05 & $-$0.15 &   0.09 &  7.00  &    1.00  &    11.71  &     700 &  0 &  7.2\\
NGC~2204    &  1.4  &  0.15 & $-$0.38 &   0.08 &  2.00  &    0.30  &    11.84  &   $-$1200 &  0 &  2.2\\
NGC~2420    &  1.6  &  0.05 & $-$0.44 &   0.06 &  2.20  &    0.30  &    10.59  &     765 &  0 &  2.8\\
NGC~6791    &  2.6  &  0.05 &   0.40 &   0.06 & 10.20   &    1.20  &     8.12  &     800 &  0 &  9.5\\
Hyades      &  0.4  &  0.05 &   0.13 &   0.06 &  0.70   &   0.10   &     8.55  &    $-$20 &  0 &  0.9\\
Praesepe    &  0.3  &  0.05 &   0.13 &   0.06 &  0.70   &   0.10   &     8.62  &      85 &  0 &  0.9\\
47~Tuc      &  2.9  &  0.05 & $-$0.70 &   0.10 & 10.90  &     1.40 &     7.40  &   $-$7400 &  0 & 12.0\\
\hline 				
\end{tabular} 
\end{minipage} 
\end{table*} 

The error bar on the individual determination has been computed by
combining in quadrature the contribution to the uncertainty  
arising from the dispersion associated to Eq.~(1), plus 
the contribution due to the [Fe/H] and $\delta(V)$
error propagation through Eq.~(1). The uncertainty in the
metallicities does not play a significant role in the error budget, due
to the weak dependence of $\log(t)$ on [Fe/H].

In Fig.~\ref{ageOC} we plot the cluster ages against their
[Fe/H]. Open squares denote the 11 calibrating clusters (10 OCs plus
47~Tuc) which cover the
entire [Fe/H] and age range spanned by the full sample, with the only
exception of Be~21, which has a metallicity $\mathrm{[Fe/H]}\sim
-1.00$, and is
therefore outside the range of validity of Eq.~(1); thus its 
age has to be treated with caution.

   \begin{figure}
     \resizebox{\hsize}{!}{\includegraphics{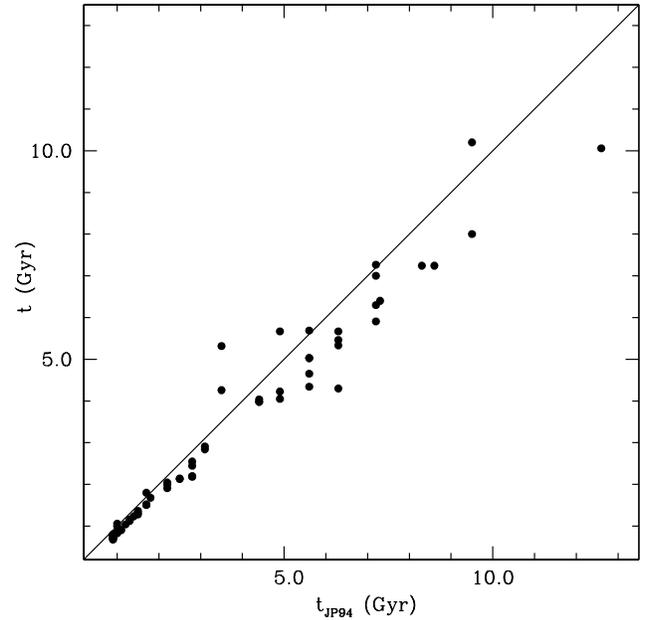}}
      \caption{Comparison between our derived ages and those given
       by JP94. The solid line denotes the 1:1 relationship.
     }
         \label{agecompOC}
   \end{figure}

Be~17 and NGC~6791 appear to be the oldest known OCs, and their ages
are formally the same, within the error bars, as the age 
of the thick disk GC 47~Tuc (see
Sect.~4 for a discussion about the comparison with GC ages).
Figure~\ref{agecompOC} compares our ages with 
the results by JP94. Our values are, with few exceptions,
systematically lower, with the oldest clusters -- Be~17 and
NGC~6791 -- having ages of about
10~Gyr, whereas the oldest cluster in JP94 is Be~17, with an
estimated age of 12.6~Gyr, about 2.5~Gyr higher than our results; our
age for Be~17 ($10.1\pm 2.8$~Gyr) is in line with the 
recent analysis by Carraro et al.~(1999) who determined 
a value of $9\pm 1$~Gyr from isochrone
fitting, although our estimate 
(based on the $\delta(V)$ provided by JP94) has a much larger error bar.
Our age for NGC~6791 -- $10.2\pm 1.2$~Gyr -- is larger
than the result by  
Chaboyer et al.~(1999) who found $8.0\pm 0.5$~Gyr from
isochrone fitting, and compatible, within the error bar, with the age
of 8--9~Gyr determined by Carraro et al.~(1999).

   \begin{figure}
     \resizebox{\hsize}{!}{\includegraphics{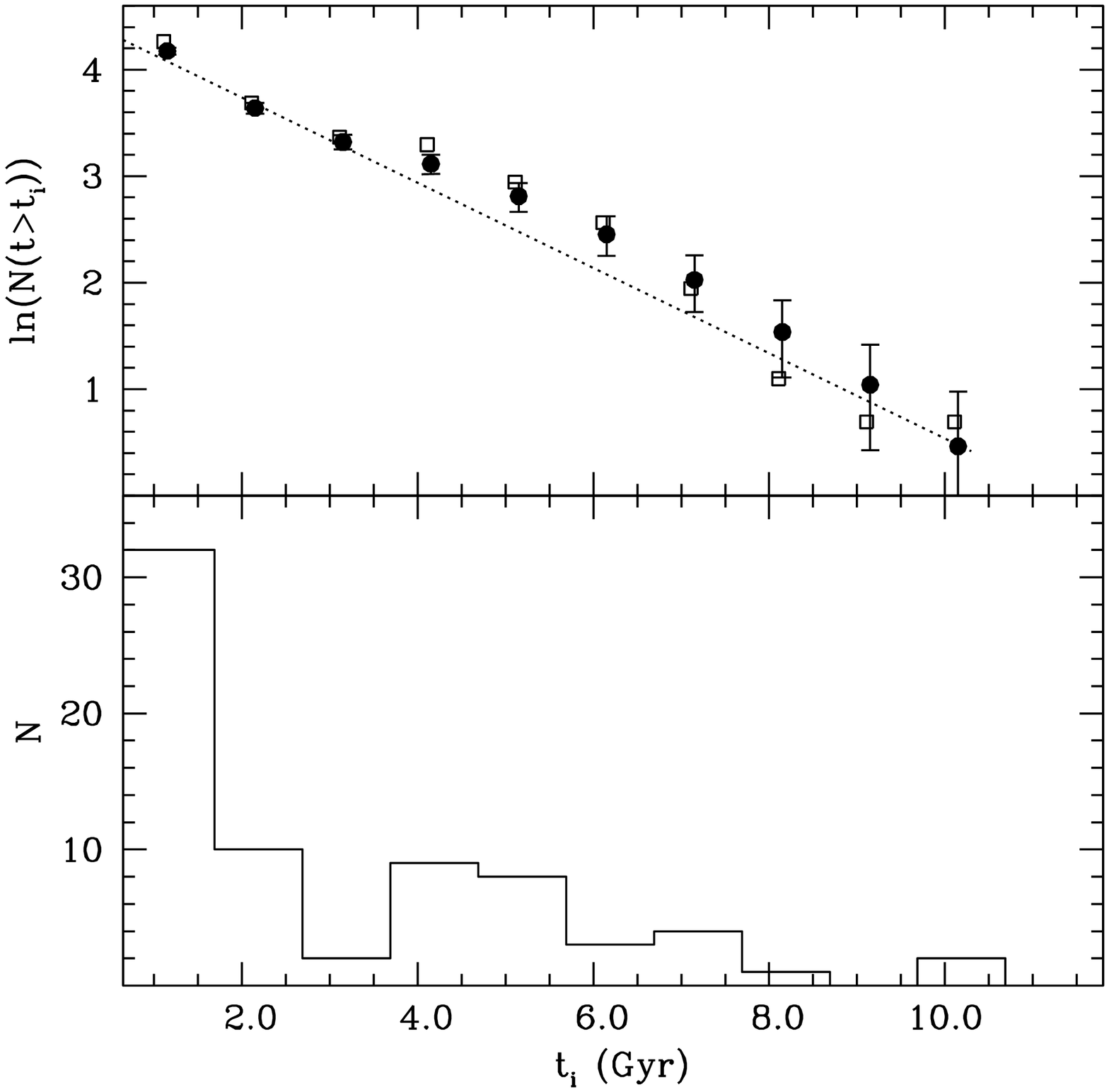}}
      \caption{Cumulative age distribution (upper panel) and
     differential age distribution (lower panel) for the 71 OCs in our sample.
     In the upper panel the age distribution from
     the Monte Carlo simulations discussed in the text (filled
     circles) is shown together with the actual cumulative distribution (open squares). 
      The line with slope corresponding to a dissolution timescale of 2.5
      Gyr is also displayed. 
      }
         \label{agehistOC}
   \end{figure}

The lower panel of Fig.~\ref{agehistOC} shows a histogram of the 
cluster ages (without the GC 47~Tuc, that in the following will
always be excluded from the analysis of the OC sample). As mentioned
before, according to JP94 the shape of this 
distribution for the whole sample should not be altered by the still
undetected old OCs. The corresponding observed cumulative function 
(i.e., number of clusters with ages larger than a given value 
$t_{i}$) is displayed in the upper panel of the same figure (open squares).

These data contain in principle important information about the
timescales of cluster destruction. In the simplest case of a
uniform formation rate and exponentially declining dissolution rate, 
the open squares in the upper panel of Fig.~\ref{agehistOC} should follow a
straight line whose slope is equal to the inverse of the dissolution timescale
(see, e.g., Janes et al.~1988). In fact, the situation
may be more complex, for the points in the cumulative age distribution
apparently do not follow a single slope; there is a change in the shape of 
the cumulative age distribution, visible 
between $\sim$4 and $\sim$6~Gyr, which appears as a 'bump'
in the differential age distribution. This 'excess' of clusters has
been already noticed by JP94, and it was located 
in the age interval between 5 and 7~Gyr on their age scale.
We have investigated further this matter by evaluating the error bars
associated to the points in our cumulative age distribution.
For this purpose, we performed an extensive 
Monte Carlo simulation by considering the
individual OC ages given in Table~1, together with their associated
errors. We then determined 10000
synthetic samples of ages for our 71 clusters. In each sample 
the individual cluster ages were randomly assigned according  
to a Gaussian distribution centred
around the values given in Table~1, with a 1$\sigma$ dispersion equal to 
the estimated individual errors. 
We then determined the cumulative age
distribution for each of the 10000 synthetic samples -- 
using the same age bins as in Fig.~\ref{agehistOC} -- 
and determined the mean number counts and associated 1$\sigma$
dispersion in each age bin (an analogous result is obtained if we use the
modal value of the number counts in each bin). 

\begin{table} 
\caption[]{Distance moduli and reddenings of the calibrating
clusters (see text for details).} 
\begin{tabular}{lcr}  \hline 
Cluster & $E(B-V)$   & $(m-M)_0$ \\  
\hline 
M~67     & 0.04  &  9.60$\pm$0.09\\ 
NGC~2477 & 0.23  & 10.74$\pm$0.08\\ 
NGC~188  & 0.09  & 11.17$\pm$0.08\\ 
NGC~7789 & 0.29  & 11.22$\pm$0.07\\ 
Be~39    & 0.11  & 12.97$\pm$0.09\\ 
NGC~2204 & 0.08  & 13.12$\pm$0.08\\ 
NGC~2420 & 0.05  & 11.94$\pm$0.07\\
NGC~6791 & 0.15  & 12.96$\pm$0.10\\ 
Hyades   & 0.00  &  3.33$\pm$0.05\\ 
Praesepe & 0.02  &  6.32$\pm$0.05\\ 
47~Tuc   & 0.04  & 13.25$\pm$0.07\\ 
\hline 
\label{OCref}
\end{tabular} 
\end{table} 

The upper panel of Fig.~\ref{agehistOC} displays the resulting synthetic cumulative age
distribution as filled circles. It is worth noticing that 
in general the actual distribution lies comfortably within the
1$\sigma$ error bars associated to the synthetic one, as expected,
since it has to correspond to one realization of our ensemble of
synthetic age distributions. We have also displayed the best fit
single slope that matches the data: 
it corresponds to a dissolution timescale of 2.5~Gyr.
Once the effect of the age
errors is taken into account, we used a $\chi^2$ test
to determine the significance of the excess of
clusters in the age range between 4 and 6 Gyr, which results to be 
at $\sim 2\sigma$ level for the whole sample of objects.

   \begin{figure}
     \resizebox{\hsize}{!}{\includegraphics{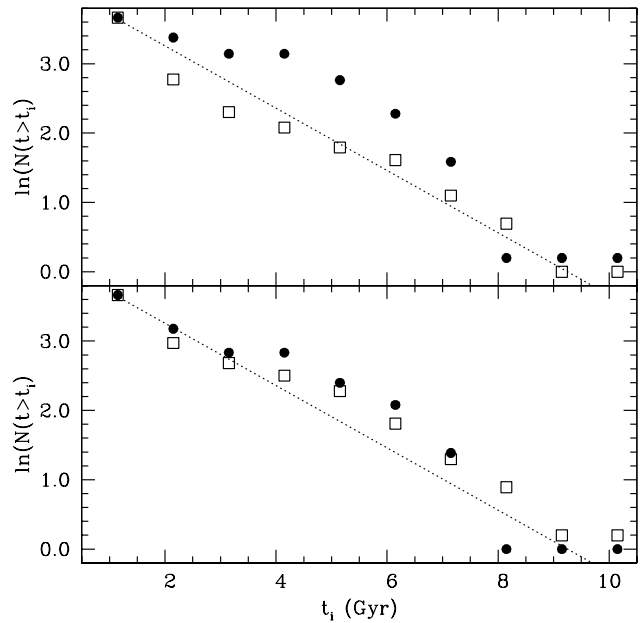}}
      \caption{Comparison of the cumulative age distributions for
      clusters located in two different ranges of height above the
      Galactic plane, $|z|$ (upper panel),  
      or galactocentric distance, $R_\mathrm{gc}$ (lower panel; see
      text for details). The 
      line with a slope corresponding to a dissolution timescale of 2.5
      Gyr is displayed in both panels.
              }
         \label{lifetimeOC}
   \end{figure}

We have also studied the dependence of the cumulative age distribution
on the cluster spatial positions.
The lower panel of Fig.~\ref{lifetimeOC} shows the 
case for the clusters located at $R_\mathrm{gc} >$10 kpc (filled
circles -- 32 objects) and at $R_\mathrm{gc} \leq$10 kpc (open squares -- 39
objects), respectively. The age distribution for clusters with 
$R_\mathrm{gc} >$10 kpc has been rescaled to have the same total number 
of objects as for $R_\mathrm{gc} \leq$10 kpc. 
The two distributions of points are very similar (the result is independent
of the definition of the $R_\mathrm{gc}$ ranges). As a guideline, the best
fit single slope corresponding to a dissolution timescale of 2.5~Gyr
is also displayed. This similarity results from the lack of correlation 
between age distribution and $R_\mathrm{gc}$ for the full sample; 
there is
no trend of age with respect to $R_\mathrm{gc}$, with a large age spread
at any value of the galactocentric distance. 
All of this, in turn, suggests that
the cluster formation and destruction processes are apparently not correlated with the
galactocentric distance.


The upper panel of Fig.~\ref{lifetimeOC} shows the cumulative age
distribution for clusters in two selected ranges 
of height above the Galactic plane $|z|$. Open squares are 
clusters with $|z|$ equal to or lower than 300 pc (39 objects), whilst filled
circles represent clusters at higher distances from the plane 
of the Galaxy (32 objects). Again, the two age distributions have been
rescaled to the same number of objects. 
It is evident that in this case the two cumulative functions
are different, i.e.\ the 
clusters closer to the plane follow a relationship much closer to the
linear slope,
corresponding to a dissolution timescale of 2.5~Gyr. 
The more distant ones show a clear excess of clusters 
in the range between 2--4 and 6~Gyr. 
This is different
from the conclusion we drew from the distribution for the whole sample
(Fig.~\ref{agehistOC}), the inconclusiveness found there might come
from mixing two different subsamples.
Analogous results are found when changing the limits to 250 pc
or 350 pc. The difference is significant (at more than 3$\sigma$
level), even considering the error
bars we obtain with a Monte Carlo simulation similar to the one discussed
for the whole sample.
This result hints at a more homogeneous creation-destruction processe for the clusters
closer to the Galactic plane, than for their more distant
counterparts. However, one has to take into account the possibility
that this difference is, at least partially, 
an artifact due to the possible incompleteness of the OC sample at low 
$|z|$ (as discussed in JP94), and/or 
to the cluster orbital motions. The analysis by Carraro \&
Chiosi~(1994b) of the orbits of 5 old OCs seem to indicate that the
observed $|z|$ for the old OCs
do not reflect their initial values, due to the rapid oscillatory
motions of the clusters across the disk.



\subsection{Age-{\rm [Fe/H]} correlation}

\begin{figure}
     \resizebox{\hsize}{!}{\includegraphics{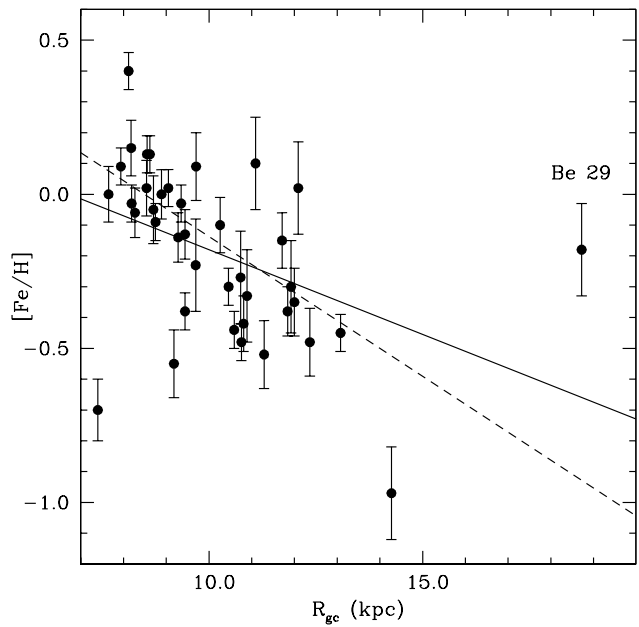}}
      \caption{Relationship between [Fe/H] and galactocentric distance
      for the 38 clusters with metallicities on the G00 scale. 
      The solid line displays the linear best-fit considering all 38
      clusters; the dashed line shows the best-fit excluding Be~29.
      }
         \label{fehrgcOC}
   \end{figure}

The determination of the age-metallicity relationship for halo and disk objects
has been the subject of numerous studies, because it poses a
constraint to the chemical evolution history of the Galaxy (e.g.,
Twarog~1980, Edvardsson et al.~1993, 
Carraro \& Chiosi~1994, Friel~1995 and references therein).
A first glance at Fig.~\ref{ageOC} does not show any trend of the
cluster age with respect to [Fe/H]. A more detailed 
analysis, however, needs to take into account the radial abundance
gradient present in the Galactic disk (e.g. Friel~1995 and references
therein). Therefore, we
have first determined the relationship -- if any -- between cluster
metallicity and galactocentric distance; we restricted our analysis to
the sample of 38 clusters with [Fe/H] on the homogeneous scale by G00
-- displayed in Fig.~\ref{fehrgcOC} -- and we fitted to the data a
linear relationship weighting the various points according to the
individual [Fe/H] error, thus obtaining

\begin{equation} 
{\rm [Fe/H]}=(-0.055\pm 0.019) \ R_\mathrm{gc} + (0.37\pm 0.20)
\label{rad1}
\end{equation} 
with a statistically significant slope, in very good agreement
with the value $-0.059\pm$0.010 dex ${\rm kpc^{-1}}$
estimated by Friel et al.~(2002) in their sample of 39 clusters.

The data displayed in Fig.~\ref{fehrgcOC} also clearly
show that the value of the slope might be affected by 
the most distant cluster in the sample, i.e. Be~29. If we exclude this
cluster the slope becomes steeper

\begin{equation} 
{\rm [Fe/H]}=(-0.097\pm 0.023) \ R_\mathrm{gc} + (0.77\pm 0.23)
\label{rad2}
\end{equation} 
but still within the range of independent determinations (see, e.g.,
the discussion by Friel et al.~2002).

We also tried to assess a possible dependence of this slope on the
cluster ages, as found by Friel et al.~(2002). As a first test, we
divided the sample into two age ranges, i.e. clusters 
with $t \leq 3$~Gyr and clusters older than 3~Gyr. For the first group
we found a gradient  $\Delta\mathrm{[Fe/H]}/\Delta
R_\mathrm{gc}=-0.13\pm0.02$~dex ${\rm kpc^{-1}}$, and
for the second one $\Delta\mathrm{[Fe/H]}/\Delta
R_\mathrm{gc}=-0.08\pm 0.03$~dex ${\rm kpc^{-1}}$. These two values are
different at about $1.5\sigma$, but the higher age group
shows a flatter gradient. This is the opposite as found by
Friel et al.~(2002) who determined a steeper gradient for clusters
older than 3~Gyr, with respect to younger objects. This could be due
to the different [Fe/H] scale they used, although their different
cluster ages may also play a role.

The precise value of the gradient for the older group is again
affected by Be~29; if we neglect this cluster 
the gradient becomes $-$0.024$\pm$0.052 dex ${\rm kpc^{-1}}$. 

This analysis clearly underscores the need to increase future 
sampling of clusters with [Fe/H] on an homogeneous scale, in order to
conclusively determine the dependence of the [Fe/H] radial gradient on age.
On the other hand, models of Galactic chemical evolution do not
provide a conclusive prediction about the age dependence of the radial
[Fe/H] gradient. As shown by Tosi~(1996), various authors predict
a [Fe/H] radial gradient which can stay constant with time, increase
or decrease (e.g. Fig.~5 in Tosi~1996).
%
%

We also checked the possible existence of an [Fe/H] gradient with
respect to the height to the Galactic plane; we considered the
[Fe/H] values the clusters
would display at the solar galactocentric distance (assumed to be
equal to 8.5 kpc), by applying the radial
gradient correction $\Delta$[Fe/H]/$\Delta
R_\mathrm{gc}$=$-$0.055 dex ${\rm kpc^{-1}}$. 
No statistical significant trend is
found. This result is the same as found by Carraro \& Chiosi~(1994a)
in a smaller sample of old OCs; according to Carraro \&
Chiosi~(1994b), this may be explained in terms of rapid oscillatory motions
of the clusters across the Galactic plane, which tend to erase any
preexisting gradient. On the other hand, the same authors conclude
that the observed radial abundance gradients should not be seriously
affected by the orbital motions, at least in the limit of the small
sample of cluster orbits (5 clusters) they analyzed.

  \begin{figure}
     \resizebox{\hsize}{!}{\includegraphics{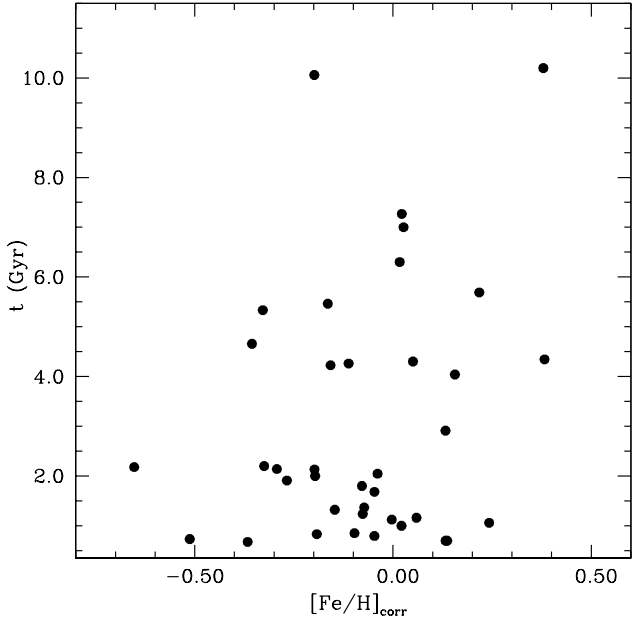}}
      \caption{Age distribution as a function of $\rm [Fe/H]_{corr}$
              ([Fe/H] corrected to the solar circle, see text for details)  
              for the 38 clusters with [Fe/H] on the G00 scale.}
         \label{agefehcorrOC}
   \end{figure}

Based on these results, we have again considered the individual 
[Fe/H] values of our 38 OCs with G00 estimates, corrected to the solar
circle ($\rm [Fe/H]_{corr}$).
The relationship between cluster age and $\rm [Fe/H]_{corr}$ 
is displayed in Fig.~\ref{agefehcorrOC}, and it
does not show any statistically significant trend between these two
quantities, just a large age spread at any metallicity. 
This is consistent with the OC results by JP94, Carraro \&
Chiosi~(1994a), Friel et al.~(2002), and also with the findings by
Edvardsson et al.~(1993) in a sample of field disk stars (although
other studies find a significant age-metallicity relationship in
field disk stars, e.g. Twarog~1980).  

\section{Comparison between GC and old OC ages}

A comparison between the ages of the oldest OCs and the GCs in the
thick disk and halo provides vital clues to the scenario for Galaxy
formation. For this comparison to be meaningful it is however necessary to ensure that
the OC and GC ages are on a consistent scale. 

In SW02 we have accurately and homogeneously 
determined the ages of a large sample of 55 GCs, using stellar models
computed with the same code and the same input physics as the models
used in this paper. The differences with respect to this work are the
age dating method and the different He enrichment law assumed in the
model computation, as discussed previously.
This latter point has been addressed by redetermining the age of the
SW02 sample using models computed with the same He enrichment law
used for the OCs. The net effect is to cause an average  
decrease by 0.7~Gyr of the GC ages with respect to SW02 results.

   \begin{figure}
     \resizebox{\hsize}{!}{\includegraphics{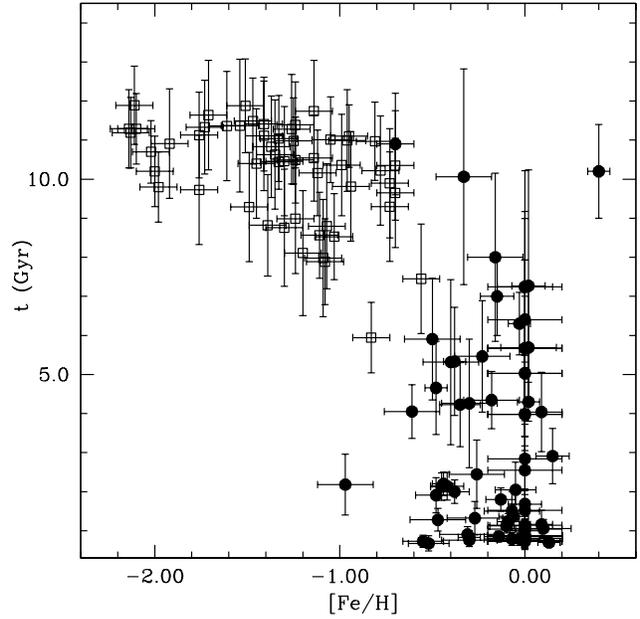}}
      \caption{Age distribution as a function of [Fe/H] for the
      clusters analyzed in this study (filled circles) and the GCs
     studied by SW02 (open squares).
              }
         \label{ageGCOC}
   \end{figure}

As for the different age dating methods applied to the two samples, 
we have one cluster -- 47~Tuc -- in common,  
whose age has been derived with both techniques.
A comparison of its age given in Table~1 
with the values from SW02 corrected for the new He abundances,
provides a difference of only 0.5~Gyr (SW02 age being lower), well
within the error bars associated to the individual determinations.

\begin{figure}
     \resizebox{\hsize}{!}{\includegraphics{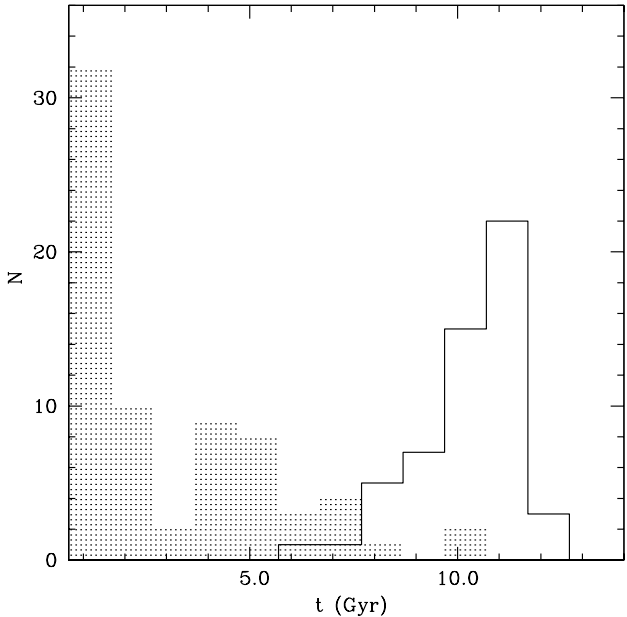}}
      \caption{Histogram of the ages of the GC sample studied by SW02,
      compared to the OC ages derived in this study (shaded histogram).
              }
         \label{histGCOC}
   \end{figure}

Figure~\ref{ageGCOC} shows the distribution of ages as a function of
the observed [Fe/H] for the SW02 sample (with metallicities according
to the Carretta \& Gratton 1997 scale) and the old OCs of Table~1,
the corresponding number counting as a function of age are displayed in 
Figure~\ref{histGCOC}. The cut-off in the age
distribution of GCs at $\sim$12~Gyr and the overlap between the tails
of the distribution of GC and OC ages is evident. The youngest GCs, supposed to
have been accreted by our Galaxy, have ages comparable to the ages of the
old OCs. This means that these accretion processes were acting well
after the formation of the Galactic disk.
NGC~6791 and Be~17, the two oldest OCs, have formally the same age as thick
disk GCs like 47~Tuc and M~71, implying an approximately 
coeval formation for both thin and thick disk.
By comparing the ages of the oldest OCs with the oldest GCs one
derives a difference  between the start of the
formation of the halo and of the thin disk of $\sim$2~Gyr .

\section{Summary and conclusions}

In this paper we have extended our previous age determinations for
GCs to old OCs belonging to the Galactic thin disk, using 
as before a morphological
age indicator. In case of the old OCs, it is the so-called 
$\delta(V)$ parameter defined by JP94.
We derived a new and homogeneous 
calibration of the $\delta(V)$--$t$--[Fe/H] relationship from a subsample of 10
clusters with accurate and deep photometry, [Fe/H]
and reddening estimates. Distances to these calibrating clusters have been
determined by means of the
MS fitting technique using field stars with {\em Hipparcos}
parallaxes, and the ages obtained from fitting the appropriate
isochrone to the absolute
brightness of the cluster turn-off region.

To obtain reliable distances and age determinations for the
calibrating clusters, a necessary prerequisite is the
use of consistent metallicity scales for both field stars and 
calibrating OCs. The metallicities for the
unevolved {\em Hipparcos} field dwarfs used in the OC age calibration were
derived by PSK02, and shown to be consistent with the G00 
metallicity scale for OCs. Comparison between the [Fe/H] estimates by 
G00 and the recent work by Friel et
al.~(2002) revealed systematic differences, the most
extreme case being NGC~6791, for which the G00 estimate is
$\mathrm{[Fe/H]} = 0.40\pm0.06$, whereas Friel et al.\ found
$0.11\pm0.10$. 
A further independent check for the internal consistency 
of our distances (and ages) is possible, 
by requiring that the distance to this cluster derived from
the MS fitting, is the same when using the $(B-V)$ or the $(V-I)$
colour. The metallicity dependence of the MS colour
is different for these two indices (see, e.g. PSK02), therefore the
consistency of the distances obtained with  $(B-V)$ and $(V-I)$ is a
good test for the adopted metallicity scales.

Keeping the cluster [Fe/H] as a free parameter, we found
that consistent distances are obtained 
-- irrespective of the choice of the cluster reddening --
only when $\mathrm{[Fe/H]}$ is 
equal to 0.4, or at least not lower than 0.3,
i.e.\ when it is homogeneous with
the scale used for the dwarfs. With our field dwarf [Fe/H] scale a
metallicity, e.g., [Fe/H]=0.2 for NGC~6791 would cause a discrepancy by 
0.11 mag between the distances inferred from the $(B-V)$ and $(V-I)$
colours.

With the $\delta(V)$--$t$--[Fe/H] relation given in Eq.~(1), we then
derived age estimates for a total of 71 OCs. Their age scale can be
merged with the one we published previously for 55 GCs (SW02), 
47~Tuc (whose age obtained in this paper agrees with the one
estimated using SW02 technique)
being the link connecting the two samples. Due to
our method, the use consistent isochrones and an homogeneous metallicity
scale, we not only obtained the first large and homogeneous sample of
OC ages, but even a reliable age scale on which both cluster types can
be placed. This allows the investigation of questions related to the
formation of the various components of the Galaxy, halo, thick and
thin disk. The bulge still awaits investigation, mainly due to the
problem of strong and differential reddening of the bulge cluster CMDs.

Using the whole GC and old OC sample (Fig.~\ref{ageGCOC}), we determine
a delay by 2.0$\pm$1.5 Gyr between the start of the halo and thin disk formation.
We estimated this value by determining the average age (with error) of
the two oldest OCs (NGC~6791 and Be~17) -- formally coeval -- 
which has been then compared with the age of the oldest metal poor GCs. 
Liu \& Chaboyer~(2000) have estimated $2.8\pm 1.6$~Gyr for this time delay,
whereas Carraro et al.~(1999) 
found the thin disk younger than the halo by
about 2--3~Gyr, an age difference shorter than the 3--5~Gyr gap determined by  
Sandage et al.~(2003).

We also find that thin and thick disk started to form approximately at the
same time, since the age of the thick disk globulars is the same,
within errors, as the age of NGC~6791 and Be~17.

The age of the oldest OCs is of the order of 10~Gyr, compatible with
that of the oldest thin disk white dwarfs as estimated from the white
dwarf luminosity function of the solar neighbourhood, which is,
according to Hansen~(1999), between
6 and 11~Gyr. This rather large age range depends on the
uncertainties in the observational data and white dwarf 
(surface and core) chemical
compositions; there are also additional uncertainties due to the 
white dwarf model physics (e.g.\ Salaris et al.~2000). 

Figure~\ref{agefehcorrOC} 
clearly demonstrates the absence of any age--metallicity relation, 
consistent with earlier results by Carraro \& Chiosi (1994a) and JP94. 
The overall slope of the relationship between [Fe/H] and $R_\mathrm{gc}$ 
is consistent with recent determinations by, e.g., Friel et al.\
(2002); however, we find a decrease of this slope for increasing 
cluster ages, which is just the opposite of the results by 
Friel et al.\ (2002). 
We do not detect any correlation between [Fe/H]
and height above the Galactic plane $|z|$, nor between age and 
$R_\mathrm{gc}$ (as in Carraro \&
Chiosi~1994 and JP94).

The cumulative age distribution for the full OC sample shows a
departure from the predictions of constant
formation rate and exponentially declining dissolution rate
(with timescale of 2.5 Gyr) at the 2$\sigma$ level. 
No correlation between the cumulative age
distribution and $R_\mathrm{gc}$ is found;
however, there is a significant excess
of clusters in the age range between 2--4 and 6 Gyr for the
population located at hig $|z|$ values, with respect to their
counterpart closer to the Galactic plane. It is not clear if this
difference is intrinsic -- i.e. related to the position of the cluster
at its birth --  or partly an artifact due to 
incompleteness of the sample (which, according to JP94, should preferentially affect
clusters with lower $|z|$) and/or to the cluster orbital motion.


\begin{acknowledgements}
This research has made use of the WEBDA data base (http://obswww.unige.ch/webda).  
SMP acknowledges financial support from PPARC.  

\end{acknowledgements}


\begin{thebibliography}{}

\bibitem[]{} Anthony-Twarog, B.J., \& Twarog, B.A. 1985, ApJ, 291, 595

\bibitem[]{} Carraro, G., \& Chiosi, C. 1994a, A\&A, 287, 761

\bibitem[]{} Carraro, G., \& Chiosi, C. 1994b, A\&A, 288, 751

\bibitem[]{} Carraro, G., Girardi, L., \& Chiosi, C. 1999, MNRAS, 309, 430

\bibitem[]{} Carretta, E., \& Gratton, R.G. 1997, A\&AS, 121, 95

\bibitem[]{} Cassisi, S., Salaris, M., \& Irwin, A.W. 2003, ApJ, 588, 862

\bibitem[]{} Chaboyer, B., Green, E.M., Liebert, J. 1999, AJ, 117, 1360

\bibitem[]{} Edvardsson, B. et al. 1993, A\&A, 275, 101

\bibitem[]{} Freeman, K., \& Bland-Hawthorn, J. 2002, ARA\&A, 40, 487
 
\bibitem[]{} Friel, E.D. 1995, ARA\&A, 33, 381

\bibitem[]{} Friel, E.D. et al. 2002, AJ, 124, 2693

\bibitem[]{} Girardi, L., Bressan, A., Bertelli, G., \&  Chiosi,
C. 2000, A\&AS, 141, 371

\bibitem[]{} Gratton, R.G. 2000, in Stellar Clusters and Associations: 
Convection, Rotation, and Dynamos, Pallavicini R., Micela G., and
Sciortino S. eds., ASP Conference Series, 198, p.225 (G00)

\bibitem[]{} Grevesse N., \& Noels, A. 1993, in Origin and Evolution
of the Elements, Prantzos N., Vangioni-Flam E., and Casse M. eds.,
Cambridge University Press, Cambridge, p.15

\bibitem[]{} Hansen, B.M.S. 1999, ApJ, 520, 680

\bibitem[]{} Janes, K. A., \& Phelps, R. L. 1994, AJ, 108, 1773 (JP94)

\bibitem[]{} Janes, K. A., Tilley, C., \& Lyng\aa, G. 1988, AJ, 95, 771

\bibitem[]{} Lejeune, T., \& Schaerer, D. 2001, A\&A, 366, 538

\bibitem[]{} Liu, W. M., \& Chaboyer, B. 2000, ApJ, 544, 818

\bibitem[]{} Mermilliod, J.-C. 1992, Bull. Inf. Centre Donnees Stellaires, Vol. 40, p.115 

\bibitem[]{} Percival, S. M., \& Salaris, M. 2003, MNRAS, 343, 539 (PS03)

\bibitem[]{} Percival, S. M., Salaris, M., \& Kilkenny, D. 2002, A\&A,
400, 541 (PSK02)

\bibitem[]{} Percival, S. M., Salaris, M., van Wyk, F., \& Kilkenny, D. 2002,
           ApJ, 573, 174 (P01)

\bibitem[]{} Phelps, R.L., \& Schick, M. 2003, AJ, 126, 265

\bibitem[]{} Phelps, R.L., Janes, K.A., \& Montgomery, K.A. 1994, AJ,
107, 1079

\bibitem[]{} Ribas, I., Jordi, C., \& Gimenez, A. 2000, MNRAS, 318, L55

\bibitem[]{} Rosenberg, A., Saviane, I., Piotto, G., \& Aparicio,
A. 1999, AJ, 118, 2306

\bibitem[]{} Salaris, M., \& Weiss, A. 1997, A\&A, 327, 107

\bibitem[]{} Salaris, M., \& Weiss, A. 1998, A\&A, 335, 943

\bibitem[]{} Salaris, M., \& Weiss, A. 2002, A\&A, 388, 492 (SW02)

\bibitem[]{} Salaris, M., Degl'Innocenti, S., \& Weiss, A. 1997, ApJ,
484, 986

\bibitem[]{} Salaris, M., Garc\'ia-Berro, E., Hernanz, M., Isern, J.,
\& Saumon, D. 2000, ApJ, 544, 1036

\bibitem[]{} Sandage, A., Lubin, L.M., \& Vandeberg, D.A. 2003, PASP,
115, 1187

\bibitem[]{} Sarajedini, A. 1999, AJ, 118, 2321

\bibitem[]{} Sarajedini, A., \& Demarque, P. 1990, ApJ, 365, 219

\bibitem[]{} Spitzer, L. 1958, ApJ, 127, 544

\bibitem[]{} Stetson, P.B., Bruntt, H., \& Grundahl, F. 2003, PASP,
115, 413

\bibitem[]{} Tosi, M. 1996, in From Stars to Galaxies: The Impact of
Stellar Physics on Galaxy Evolution,
Leitherer C., Fritze-von-Alvensleben U., and Huchra J. eds., 
ASP Conference Series, 98, p.299

\bibitem[]{} Twarog, B. A. 1980, ApJ, 242, 242

\bibitem[]{} Vandenberg, D. A., Bolte, M., \& Stetson, P. B. 1990, AJ,
100, 445

 
\end{thebibliography}
\end{document}